\documentclass[conference]{IEEEtran}
\usepackage[utf8]{inputenc}
\usepackage{graphicx}

\graphicspath{{imgs/}}

\usepackage{amsmath,amssymb,amsfonts,stmaryrd}
\usepackage{graphicx}
\usepackage{textcomp}
\usepackage[table]{xcolor}
\usepackage{xcolor}
\usepackage{balance}
\usepackage{subcaption}

\usepackage{algorithm}
\usepackage[noend]{algpseudocode}

\usepackage{hyperref}

\usepackage{dblfloatfix}

\usepackage{soul}

\usepackage{biblatex}
\algnewcommand\algorithmicforeach{\textbf{for each}}
\algdef{S}[FOR]{ForEach}[1]{\algorithmicforeach\ #1\ \algorithmicdo}

\author{
\IEEEauthorblockN{Robert Šajina\IEEEauthorrefmark{1}}
\IEEEauthorblockA{\textit{Faculty of Informatics}\\
\textit{Juraj Dobrila University of Pula}\\
Pula, Croatia\\
robert.sajina@unipu.hr\\
\IEEEauthorrefmark{1}Corresponding author
}
\and
\IEEEauthorblockN{Ivo Ipšić}
\IEEEauthorblockA{\textit{Faculty of Informatics and Digital Technologies}\\
\textit{University of Rijeka}\\
Rijeka, Croatia\\
ivoi@uniri.hr
}
}

\title{Sequence-to-sequence models in peer-to-peer learning: A practical application}

\begin{document}

\maketitle

\begin{abstract}

This paper explores the applicability of sequence-to-sequence (Seq2Seq) models based on LSTM units for Automatic Speech Recognition (ASR) task within peer-to-peer learning environments.
Leveraging two distinct peer-to-peer learning methods, the study simulates the learning process of agents and evaluates their performance in ASR task using two different ASR datasets.
In a centralized training setting, utilizing a scaled-down variant of the Deep Speech 2 model, a single model achieved a Word Error Rate (WER) of 84\% when trained on the UserLibri dataset, and 38\% when trained on the LJ Speech dataset.
Conversely, in a peer-to-peer learning scenario involving 55 agents, the WER ranged from 87\% to 92\% for the UserLibri dataset, and from 52\% to 56\% for the LJ Speech dataset.
The findings demonstrate the feasibility of employing Seq2Seq models in decentralized settings, albeit with slightly higher Word Error Rates (WER) compared to centralized training methods.

\end{abstract}

\section{Introduction}

Sequence-to-sequence (Seq2Seq) models have emerged as a powerful framework in various natural language processing (NLP) and sequence generation tasks, ranging from machine translation \cite{Diyah2021}, text summarization \cite{Tian2021} to even automatic speech recognition (ASR) \cite{Li2021}.
However, their application in peer-to-peer deep learning scenarios \cite{Blot2016}, even within the realm of Federated Learning (FL) \cite{BrendanMcMahan2017}, remains relatively under explored.
Peer-to-peer and FL environments are comprised of agents, each holding local private data and preforming local computations, with the primary mode of interaction being the sharing of models with outside entities or other agents.
Agent's local data is considered as private and must never leave the agent, therefore only the machine learning models are exchanged with other participants in the environment.
In the context of Federated Learning (FL), agents disseminate their trained models solely with a central server, whereas in peer-to-peer settings, agents engage in direct model exchange among themselves, thereby rendering peer-to-peer environments notably more complex scenarios.

An agent, which may be any edge device such as phone or laptop, owns local data which is most likely generated by the agent or the user using the device. 
When considering the applications of peer-to-peer learning, the availability of local data must also be considered.
In this context, local textual exchanges or user-generated written content emerge as potential dataset that can be utilized for training neural network models tailored to tasks like next-word prediction \cite{Sajina2023}.
In the context of Seq2Seq model applications, an Automatic Speech Recognition (ASR) task presents itself as a viable option.
Users can contribute to the generation of data by orally presenting specific paragraphs or sentences, therefore creating a dataset of audio clips and corresponding textual representations which can be utilized to train a local Seq2Seq model.
The purpose of this study is to investigate if such application of the Seq2Seq model is viable peer-to-peer environment.
Reproducible code is available on a publicly available repository: \url{https://github.com/rosaj/p2p_bn}.
\section{Background}

\subsection{Peer-to-peer learning}
In peer-to-peer learning, agents exchange their models following a network topology which is commonly predetermined \cite{Assran2019}, however, the peer connection between agents can also be established autonomously based on agents' preferences \cite{ZexiLi2022}.
Since the topology often dictates the number of neighbors, it is important to note that the communication can be directed or undirected.
In an example of undirected ring topology, each agent has two neighbors, while in the directed ring topology, each agent has only one neighbor.

Each agent's goal is to train its local model by leveraging its local dataset and information received from its neighbors to minimize its average loss function.
An agent uses this local data to train a local model by calculating mini-batch gradient and updating its local model for predetermined number of batch iterations.
The local training step is followed by a communication step in which agents exchange information. 
Scenarios involving multiple agent learning and interacting with eachother are often simulated in memory in a computer, in parallel, i.e., each agent performs training and communication steps in a loop.
Different approaches include different message content exchanged between agents; commonly, an agent must send its message to all its out-neighbors and receive all messages from its in-neighbors.
A loss of a message in this phase may stall the overall learning process since an agent may forever wait for a message that never arrives.

In a synchronous approach, a synchronization barrier is implemented, mandating that agents refrain from proceeding with their learning processes until all messages have been both transmitted and received \cite{Guo2020}.
Conversely, an asynchronous approach permits agents to engage in communication and learning activities at their discretion, without significant constraints \cite{Robert2020}.
Furthermore, variations in the aggregation methodology may arise, with certain techniques involving the aggregation of all received models by computing the average of all model parameters, inclusive of local model parameters \cite{Assran2019}, while in alternative methods, an agent may conduct averaging operations upon each received model in conjunction with its own local model subsequent to reception \cite{Sajina2023}.


\subsection{Sequence-to-sequence models for ASR}
\begin{figure}[htbp]
\centering
\includegraphics[width=1\linewidth]{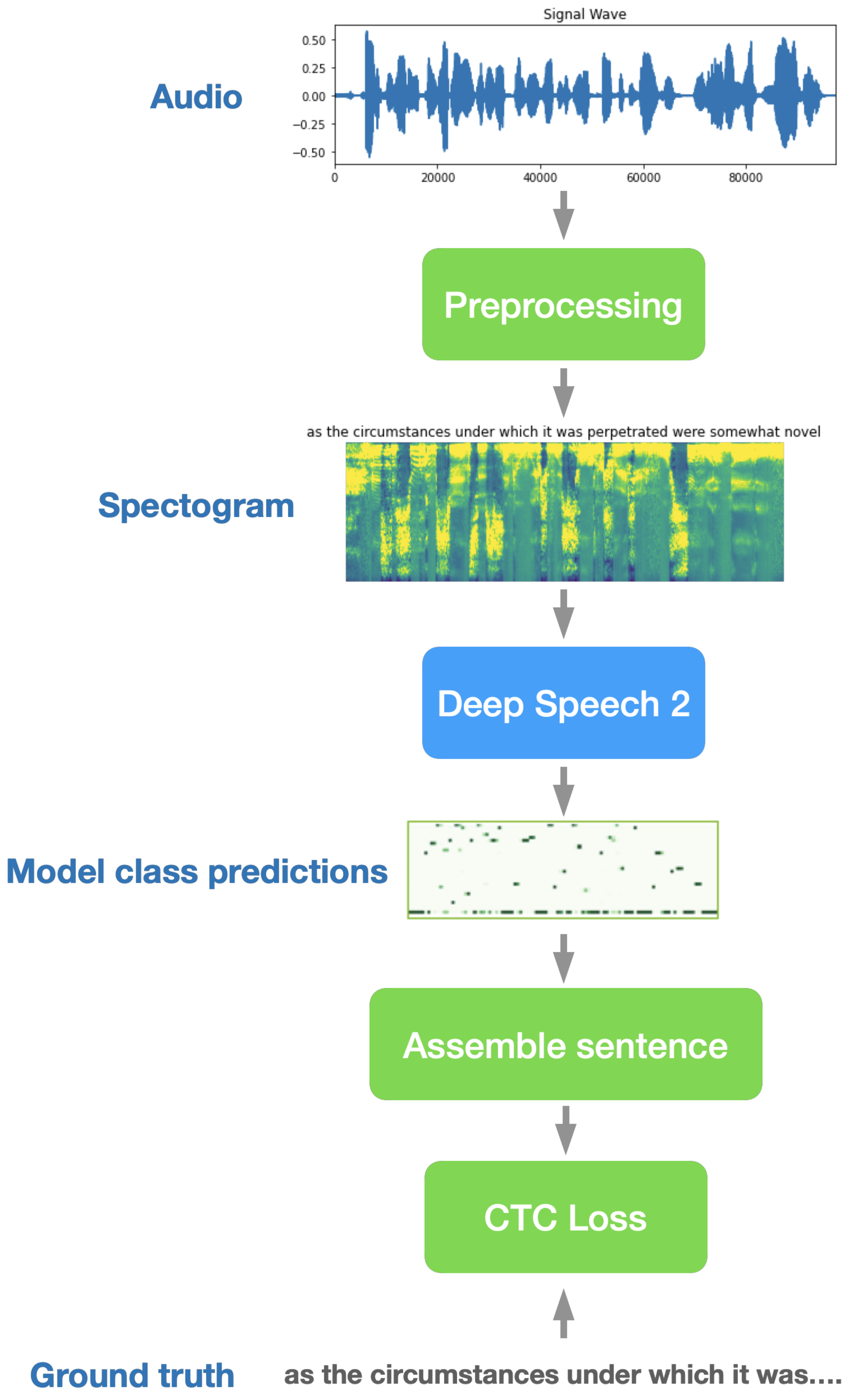}
  \caption{Data processing and training scheme.
    \label{fig:processing}
  }
\end{figure}

Seq2Seq models consist of two main components: an encoder and a decoder.
In the context of ASR, the encoder processes input speech signals, typically represented as spectrograms or other time-frequency representations, into a fixed-dimensional vector representation, capturing relevant features of the input audio.
The decoder, often implemented as a recurrent neural network (RNN) or transformer architecture, then generates the corresponding text output based on the encoded information.
The most known model architectures regarding the ASR tasks are the \textit{Listen, Attend and Spell} (LAS) \cite{LAS2015} and Deep Speech 2 \cite{DeepSpeech2}.
In this study, Deep Speech 2 model variant will be used to investigate the applicability of the Seq2Seq models in peer-to-peer environments for the ASR task.
In conventional training process (depicted in Figure \ref{fig:processing}, the audio clip is first converted to a spectrogram which is essentially an image, which is then used as an input to the model.
The model outputs a sequence of token probabilities which is then used to assemble the output sentence. 
Based on specific needs, the output may be individual characters or words.
Once the sentence is assembled, Connectionist Temporal Classification (CTC) \cite{Alex2006CTC} loss metric is used to evaluate the similarity of the model output as compared to ground truth text that was spoken in the audio clip.

The primary evaluation metric utilized in the assessment of automatic speech recognition systems is the Word Error Rate (WER).
WER quantifies the minimum number of operations, including substitutions, deletions, and insertions, necessary to transform the system's transcription (prediction) into the reference transcription (ground truth), divided by the total number of words in the reference.
WER is bounded between 0 and infinity, where a lower value indicates better performance.
Often expressed as a percentage, WER is commonly calculated by multiplying the raw score by 100.
For instance, a WER of 0.15 is equivalently represented as 15\%. 
Extraction of words from a trained model necessitates the utilization of a decoder, which translates a probability distribution over characters into textual output.
Two primary types of decoders are typically employed with Connectionist Temporal Classification (CTC)-based models: the greedy decoder and the beam search decoder with language model re-scoring.
The greedy decoder selects the most probable character at each time step, facilitating rapid inference and generating transcripts closely resembling the original pronunciation.
However, this approach may introduce numerous minor misspelling errors, which, given the nature of the WER metric, render entire words incorrect with even a single character discrepancy.
Conversely, the beam search decoder with language model re-scoring explores multiple potential decodings concurrently, assigning higher scores to more probable N-grams based on a specified language model.
Integration of the language model aids in rectifying misspelling errors. 
Nevertheless, this approach is notably slower in comparison to the greedy decoder.
For simplicity and speed, this study utilizes the greedy decoder.
\section{Related work}

Research in the domain of Seq2Seq models for Automatic Speech Recognition (ASR) has been extensively explored and documented within the context of training a single centralized model \cite{Fan2020, Li2021}.
However, the applicability of these models within decentralized systems remains an area requiring further investigation.
In the realm of centralized distributed learning, such as Federated Learning, numerous studies have emerged involving sequence-to-sequence models across various natural language processing (NLP) tasks.

For instance, Lin \textit{et al.} \cite{Lin2022} developed FedNLP, a benchmarking framework tailored for evaluating Federated Learning methods on common NLP tasks. 
his framework utilizes Transformer-based language models for tasks like text classification, sequence tagging, question answering, and sequence-to-sequence generation.
Similarly, Lu \textit{et al.} \cite{Lu2021} proposed the Federated Natural Language Generation (FedNLG) framework, which facilitates the learning of personalized representations from distributed datasets across devices.
FedNLG enables the implementation of personalized dialogue systems by pre-training standard neural conversational models over large dialogue corpora and fine-tuning model parameters and persona embeddings in a federated manner.

Furthermore, recent research by Nguyen \textit{et al.} \cite{Nguyen2023} specifically addresses the ASR task within the context of Federated Learning.
Their study employs the Wav2vec 2.0 model \cite{Baevski2020}, which features a Transformer-influenced architecture, demonstrating the feasibility and effectiveness of leveraging advanced neural network architectures for speech recognition tasks within decentralized learning frameworks.

Related work does not address the applicability of Seq2Seq models based on LSTM cells for the ASR task, especially within the peer-to-peer environments.
Several potential problems may arise from the architecture comprised of LSTMs, as the vanishing or exploding gradients problems associated with RNN cells.
This problems may especially be highlighted due to the nature of local datasets comprised on agents which may contain small amount of data and differ from agent to agent \cite{Ma2022}.
This work will analyze the applicability of Seq2Seq model based on architecture utilizing LSTM cells for agents collaborating in peer-to-peer environments.

The existing literature lacks a comprehensive exploration of the applicability of Seq2Seq models based on LSTM cells for the ASR task, particularly within peer-to-peer environments.
Such architectures pose several potential challenges, including issues related to vanishing or exploding gradients commonly associated with recurrent neural network (RNN) cells.
These challenges may be exacerbated by the nature of local datasets held by individual agents in peer-to-peer settings, which may vary in size and composition \cite{Ma2022}.
By investigating the performance of such models under two different communication and aggregation techniques, this research seeks to elucidate the potential benefits and limitations of employing LSTM-based Seq2Seq models in decentralized learning settings.
Through empirical evaluations and experimental validations, this work aims to provide insights into the feasibility and effectiveness of utilizing LSTM-based Seq2Seq models for ASR tasks in distributed peer-to-peer environments.
\section{ASR in peer-to-peer learning}

\begin{figure*}[htbp]
\centering
\includegraphics[width=1\linewidth]{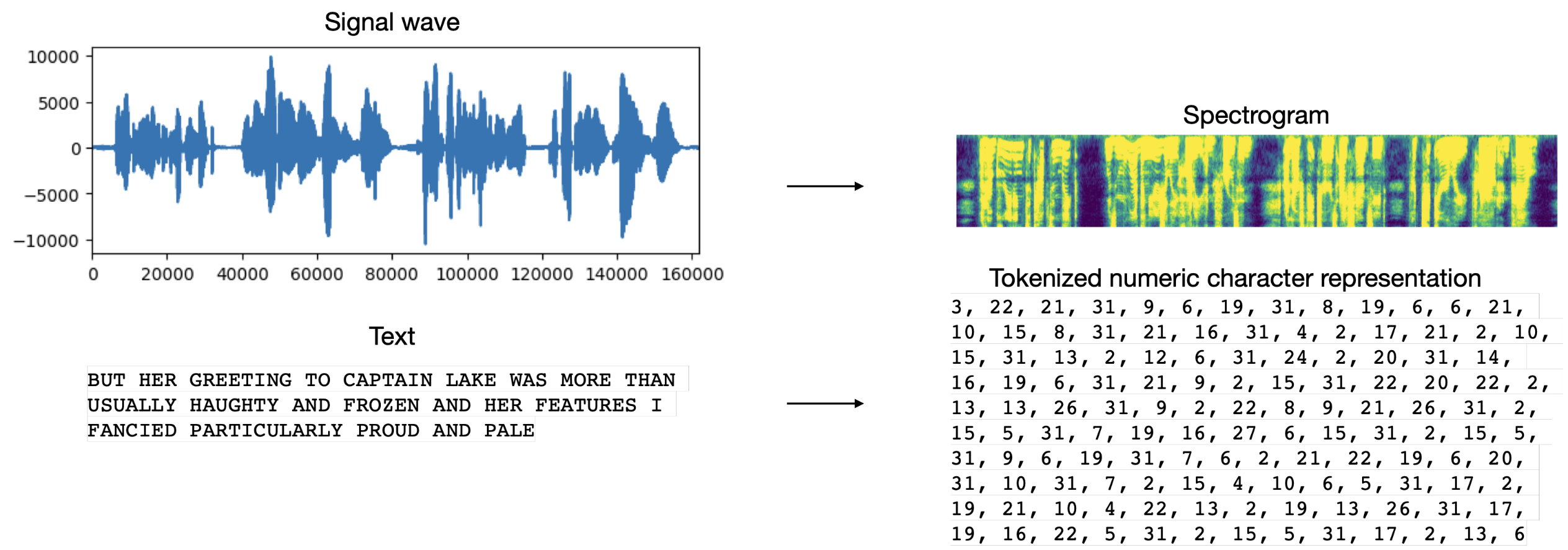}
  \caption{An example of raw and processed audio-text pair from the UserLibri dataset.
    \label{fig:userlibri_example}
  }
\end{figure*}

\noindent
\textbf{Dataset.}
UserLibri \cite{UserLibri} dataset will be used in the experiments.
This UserLibri dataset is a re-formatted version of the LibriSpeech data \cite{Librispeech} that is derived from English audiobooks and contains 1000 hours of speech sampled at 16 kHz.
An example of audio-text pair from the UserLibri dataset is shown in Figure \ref{fig:userlibri_example}.
In UserLibri, the data is reorganized into individual \textit{user} datasets consisting of paired audio-transcript examples and domain-matching text-only data for each user.
Data segregated in this manner allows for a simulation of realistic scenario with user-specific audio clips.
The UserLibri dataset contains 55 unique users (as part of the test-clean split), with average of 47.1 audio clips per user.
The duration of audio samples ranges from 1 to 35 seconds, with an average length of 7.5 seconds, resulting in approximately 5 hours of audio recording in total.
Within these audio clips, the number of spoken words spans from 1 to 96, with an average of 20 words per clip.
Such small amount of local data may cause problems during local training which may lead to poor model performance.
Threfore, an additional LJ Speech dataset \cite{ljspeech17} will be used in additional experiments to confirm any findings from the experiments regarding the UserLibri dataset.

The LJ Speech dataset consists of 13,100 short audio clips with accompanying transcriptions for clip.
The audio recordings sourced from the LJ Speech dataset exhibit varying durations, spanning from 1 to 10 seconds, with an average duration of 6.5 seconds per clip, resulting in around 24 hours of audio recordings in total.
Within these recordings, the number of spoken words ranges from 1 to 39, with an average of 17 words per clip.
Both datasets were split in a 70\%-30\% for training and validation purposes.

\begin{figure*}[htbp!]
\centering
\begin{subfigure}{0.49\linewidth}
    \includegraphics[width=\linewidth]{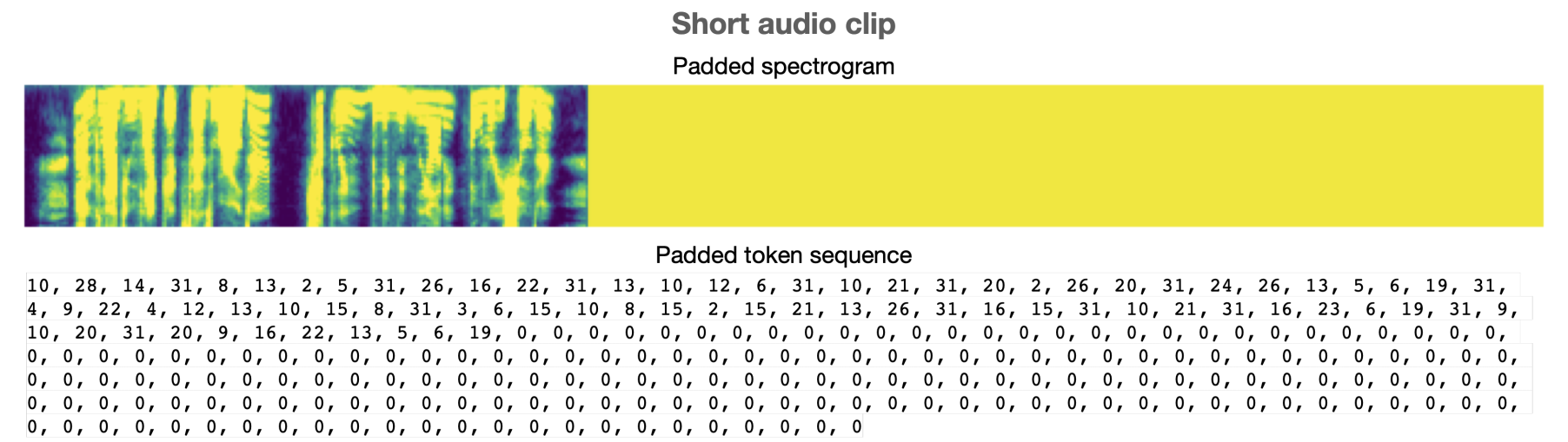}
\end{subfigure}
\begin{subfigure}{0.49\linewidth}
    \includegraphics[width=\linewidth]{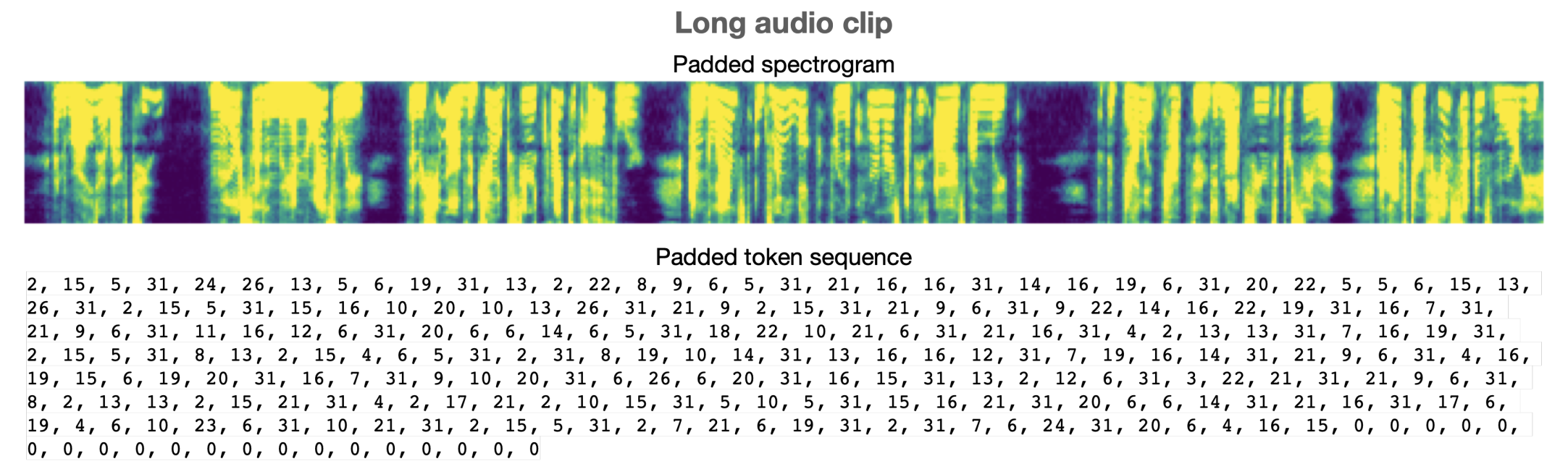}
\end{subfigure}
\caption{Examples of short audio-text pair with added audio and text padding, compared to a long audio-text pair.}
\label{fig:padding}
\end{figure*}

To ensure uniformity in input dimensions for training, spectrograms and character tokens underwent a padding process, aligning them to a consistent length.
Specifically, each spectrogram was extended or truncated to a fixed length of 2048 time steps, while character tokens were padded or trimmed to 256 characters.
Spectrograms and token sequences were padded with zeros where necessary. An example of short and long audio-text sequence in Figure \ref{fig:padding}.

\medskip
\noindent
\textbf{Model.}
Model architecture consists of two blocks of 2D Convolutional layers that are followed by a Batch Normalization layer and ReLu activation function.
A single RNN GRU layer with 512 units is used to constrain the number of model parameters. GRU layer is followed by a Fully connected layer, Dropout layer, and the final Fully connected output layer.
The resulting model is comprised of 7.7M model parameters which is of acceptable size for edge devices, and satisfactory for this use case since the goal is to evaluate the applicability of Seq2Seq models in peer-to-peer environments, rather than achiveing state-of-the-art results.
To reduce memory consumption and improve the training process, batch size is set to 8.
Adam \cite{kingma2017adam} optimizer with a fixed learning rate of 0.0001 is used in all experiments.
Model predicts the probabilities of each individual character of English alphabet and uses the greedy decoder to select the most probable character at each time step.

\medskip
\noindent
\textbf{Methodology.}
The learning process of agents was emulated in memory using two distinct peer-to-peer learning methodologies: Pull-gossip \cite{Jin2016} and P2P-BN \cite{Sajina2023}.
This simulation engenders a cyclical process wherein each local training step is succeeded by a communication step, with these phases iteratively recurring.
A directed \textit{sparse} network topology with three peer connections was utilized for both methods.
In the Pull-gossip approach, upon pulling models from its peers, an agent forms a new model by aggregating all received models using a weighted summation mechanism: $x_i = \sum_{\{w_{ij}>0\}} w_{ij}x_j$, where $x_i$ represents the new model of agent $i$, $x_j$ denotes the received model from peer $j$, and $w_{ij}$ signifies the weight associated with the connection between agents $i$ and $j$.
Conversely, in the P2P-BN method, upon receiving a model from a peer, an agent constructs a new model by averaging the received model with its local model: $x_i = \frac{x_i+x_j}{2}$.

Agent's local test accuracy is measured after performing local SGD updates (i.e. an epoch).
As comparison, results of a centrally trained model on pooled data are also analyzed.
Experiments regarding UserLibri dataset utilized all 55 user dataset in a 55 agent environment, while the data from LJ Speech was uniformly divided across 55 agents in the experiments utilizing the LJ Speech dataset.
To assess the performance of agents' models, the average model CTC loss and Word Error Rate (WER) metrics will be examined.
The computation of average local model performance across all agents using their respective local datasets serves as a pivotal evaluation metric within peer-to-peer environments \cite{Mills2022, Sajina2023, SajinaMT2023}.

\section{Experiments}

\subsection{Experiments on UserLibri dataset}
To establish a baseline for the optimal model performance, a central model is trained using aggregated data from all 55 users within the UserLibri dataset.
Since batch size of 8 used, this resulted in 226 training batches which showed to be insufficient to train a good model. 
The onset of overfitting was observed approximately at the 10th epoch, while the lowest WER achieved was 84.34\%, with character error rate (CER) at 39.42\%.
These suboptimal outcomes indicate that the dataset's size may be insufficient to adequately train the model, thereby limiting its capacity to achieve satisfactory performance levels.
For reference, initial experiments conducted on the UserLibri dataset obtained average per-user WER value of around 2.5\% \cite{UserLibri} by using a 86M parameter Conformer Hybrid Autoregressive Transducer (HAT) \cite{Variani2020HybridAT} model.

The following are some examples of target (correct) and (model) prediction sentences:

\noindent
\begin{itemize}
    \item Example 1
        \begin{itemize}
        \item  \textbf{Target}: A ROBBER VIKING SAID THE KING AND SCOWLED AT ME
        \item \textbf{Prediction}: A R BER FI GENGC AD THE CING AND Y SCOULD IT ME
        \end{itemize}
        
    \item Example 2
    \begin{itemize}
        \item  \textbf{Target}: EDISON HELD THAT THE ELECTRICITY SOLD MUST BE MEASURED JUST LIKE GAS OR WATER AND HE PROCEEDED TO DEVELOP A METER
        \item \textbf{Prediction}: ET ISSUNHO THET HE ULECTRISINTE SID MUST BEIN MISUR JUS LI GIS OR OT ER AND YH PRESUTOD TO DEVU MITER
    \end{itemize}
        
    \item Example 3
    \begin{itemize}
        \item  \textbf{Target}: WHETHER OR NOT THIS PRINCIPLE IS LIABLE TO EXCEPTIONS EVERYONE WOULD AGREE THAT IS HAS A BROAD MEASURE OF TRUTH THOUGH THE WORD EXACTLY MIGHT SEEM AN OVERSTATEMENT AND IT MIGHT SEEM MORE CORRECT TO SAY THAT IDEAS APPROXIMATELY REPRESENT IMPRESSIO
        \item \textbf{Prediction}: WTHER NOT THESPRENCSPUD IS LBLE TEXKCEPIOANS EVERE ONE WUDA GRETHEK IS HAS ABROUDNMUSURE TRTE BHE HEARED EXSARCCLAY MENEN OVERSSTEGMKT AND IT MASENG MOR CURECT TISSAY THAT IEAD DIS UPROCSIN  YD REPRENTUND PTIONS
    \end{itemize}
\end{itemize}

As stated, in the peer-to-peer environment, each user autonomously trains their respective model utilizing locally available data sourced from the UserLibri dataset, subsequently exchanging their locally trained model with peers.
Evidently, the peer-to-peer setting necessitates significantly more training iterations to attain performance levels comparable to those of centrally trained models. 
Despite yielding slightly higher WER, with Pull-gossip method achieving 92\% and P2P-BN method achieving 87\%, the findings indicate that Seq2Seq models can indeed be applied to ASR tasks within peer-to-peer environments. 


\subsection{Experiments on LJ Speech dataset}
A central model was trained on all training LJ Speech dataset to establish a baseline result. 
The model was able to substantially better results as compared to the previous model trained on UserLibri dataset, mainly because of the larger number of samples.
While the model exhibited some form of overfitting, observed by the divergence between training and validation loss, the model still managed to achieve quite low WER value, around 38.61\% with CER at 10.59\%.
For reference, recent study demonstrated that standard DeepSpeech2 architecture achieved a WER value of 0.1\% \cite{Lau2023}.

\noindent
The following are some examples of target (correct) and (model) prediction sentences:

\begin{itemize}
    \item Example 1
        \begin{itemize}
        \item  \textbf{Target}: NO NIGHT CLUBS OR BOWLING ALLEYS NO PLACES OF RECREATION EXCEPT THE TRADE UNION DANCES I HAVE HAD ENOUGH
        \item \textbf{Prediction}: NOW NHIGH KCLOBS ER BOULLIG ALLIYS NO PLACES OFREACRIATION EXEPT THE TRA UGIN DANCES I HAVD HA A
        \end{itemize}
        
    \item Example 2
    \begin{itemize}
        \item  \textbf{Target}: TO AN INFERENCE THAT THE VIOLATION OF THE REGULATION HAD CONTRIBUTED TO THE TRAGIC EVENTS OF NOVEMBER TWENTYTWO
        \item \textbf{Prediction}: TO AN INFRIND S THAT THE VIOLATION OF THE REGULATION AD CONTRIBIUTED TO THE TRAGICGAVANS OF NOVEMBER TWENTYTWO
    \end{itemize}
        
    \item Example 3
    \begin{itemize}
        \item  \textbf{Target}: UNTIL APRIL NINETEEN SIXTY FBI ACTIVITY CONSISTED OF PLACING IN OSWALD'S FILE
        \item \textbf{Prediction}: UNTIL APRIL NINETEEN SIXTY FBI ACTIVITY CONSISTED OF PLACING IN OSWALD'S FIL
    \end{itemize}
\end{itemize}

In the peer-to-peer scenario, the LJ Speech dataset was uniformly partitioned among the 55 agents, resulting in an allocation of 166 training examples per agent. 
This allocation represents a considerable increase, exceeding threefold, in training data availability compared to the experiment conducted with the UserLibri dataset.
Consistent with prior experiments, peer-to-peer learning necessitated a notable increase in training iterations to achieve model convergence.
However, it is noteworthy that the number of training iterations until convergence remained comparable to previous experiments, suggesting that increase of local data solely enhanced model performance without necessitating a commensurate increase in communication or training iterations. 
Within this experiment, agents attained a WER value of 56\% for the Pull-gossip and 52\% for the P2P-BN method.



\section{Conclusion}

In conclusion, this discussion has shed light on the utilization of Seq2Seq models in both centralized and peer-to-peer learning environments for Automatic Speech Recognition (ASR) tasks.
While centralized training on pooled data from all users demonstrated more efficient convergence and lower Word Error Rates (WER), peer-to-peer learning, where each agent independently trains its own model using local data, showcased promising potential despite requiring more training iterations to achieve comparable results.
These findings underscore the applicability of Seq2Seq models in peer-to-peer settings.
However, a critical observation gleaned from the experiments is the pivotal role played by the availability and richness of local training data across decentralized agents.
The correlation between larger local datasets and lower Word Error Rates (WER) on the LJ Speech dataset underscores the importance of data quantity in achieving optimal performance with Seq2Seq models. However, it's crucial not to assume this constraint for decentralized agents, as they may not have access to substantial amounts of local data. Therefore, future research endeavors should prioritize the exploration and development of methods that enable robust learning performance for Seq2Seq models, even in scenarios where agents have access to only small quantities of data
Additionally, future research endeavors should focus on addressing challenges such as slow convergence rates in peer-to-peer learning and devising techniques to expedite the learning process, thereby facilitating more efficient and scalable deployment of Seq2Seq models in decentralized settings.

\printbibliography
\end{document}